\ifx\destination\undefined
  \def\destination{aanda}  
\fi
\def\arxiv{arxiv}
\def\aanda{aanda}
\def\publisher{publisher}
\RequirePackage{snapshot}
\PassOptionsToPackage{dvipsnames}{xcolor}
\PassOptionsToPackage{colorlinks=True,allcolors=RoyalBlue}{hyperlink}

\ifx\destination\arxiv
  \documentclass[ut8,T1]{article}
  \pdfoutput=1	
  \usepackage{aux/preprint}
  \usepackage{natbib}
  \usepackage{authblk}
  \let\address\affil
\fi
\ifx\destination\aanda
  \documentclass[ut8,T1]{aa}
  \usepackage{natbib}
  \bibpunct{(}{)}{;}{a}{}{,} 
\fi
\ifx\destination\icarus
  \documentclass[preprint,3p,authoryear,lefttitle,utf8,T1]{elsarticle}
  \journal{Icarus}
\fi

\usepackage[utf8]{inputenc}
\usepackage[T1]{fontenc}
\usepackage{xspace}
\usepackage{ulem}
\usepackage{listings}
\usepackage[dvipsnames]{xcolor}
\usepackage{dirtytalk}

\definecolor{codegreen}{rgb}{0,0.6,0}
\definecolor{codegray}{rgb}{0.5,0.5,0.5}
\definecolor{codepurple}{rgb}{0.58,0,0.82}
\lstdefinestyle{mystyle}{
    commentstyle=\color{codegray},
    keywordstyle=\color{magenta},
    numberstyle=\tiny\color{codegray},
    stringstyle=\color{codegreen},
    basicstyle=\ttfamily\footnotesize,
    breakatwhitespace=false,
    breaklines=true,
    captionpos=b,
    keepspaces=true,
    showspaces=false,
    showstringspaces=false,
    showtabs=false,
    tabsize=2
}
\usepackage{siunitx}
\DeclareSIPrefix\micro{\text{\textmu}}{-3}
\usepackage{aas_macros} 

\usepackage{graphicx}
\usepackage{here}
\usepackage{tikz}
\usepackage{pgf}
\newcommand{\mathdefault}[1][]{}
\usepackage{import}  
\graphicspath{{gfx/}}

\usepackage[title]{appendix}
\usepackage{booktabs}
\usepackage[skip=4pt]{caption}
\usepackage[switch]{lineno}

\usepackage{url}
\usepackage[colorlinks=True,
            linkcolor=gray,         
            citecolor=RoyalBlue,    
            filecolor=gray,         
            urlcolor=gray           
           ]{hyperref}

\usepackage[nameinlink,capitalize]{cleveref}
\crefformat{equation}{Eq.~#2#1#3}
\crefname{section}{Sect.}{Sects.} 

\newcommand{\numb}[1]{\textcolor{orange}{#1}}
\renewcommand{\numb}[1]{#1}

\newcommand{\nuna}[2]{(\num{#1}) \textit{#2}}  
\newcommand{\miriade}{\texttt{Miriade}\xspace}
\newcommand{\ssodnet}{\texttt{SsODNet}\xspace}
\newcommand{\rocks}{\texttt{rocks}\xspace}
\newcommand{\topcat}{\texttt{TOPCAT}\xspace}
\newcommand{\astropy}{\texttt{astropy}\xspace}
\newcommand{\sbpy}{\texttt{sbpy}\xspace}
\newcommand{\astorb}{\texttt{astorb}\xspace}
\newcommand{\gaia}{\textsl{Gaia}\xspace}
\newcommand{\euclid}{\textsl{Euclid}\xspace}
\newcommand{\tess}{\textsl{TESS}\xspace}
\newcommand{\kepler}{\textsl{Kepler/K2}\xspace}
\newcommand{\juice}{\textsl{JUICE}\xspace}
\newcommand{\fink}{\textsc{Fink}\xspace}
\newcommand{\apisso}{\textsc{sso}\xspace}
\newcommand{\ssoft}{\textsc{ssoft}\xspace}
\newcommand{\atlas}{\textsc{ATLAS}\xspace}
\newcommand{\ztf}{\textsc{ZTF}\xspace}
\newcommand{\sdss}{\textsc{SDSS}\xspace}
\newcommand{\sm}{\textsc{SkyMapper}\xspace}
\newcommand{\ps}{\textsc{PanSTARRS}\xspace}
\newcommand{\funcf}{\ensuremath{f(r,\Delta)}\xspace}
\newcommand{\funcg}{\ensuremath{g(\gamma)}\xspace}
\newcommand{\funcs}{\ensuremath{s(\alpha,\delta)}\xspace}
\newcommand{\hg}{\texttt{HG}\xspace}
\newcommand{\hgs}{\texttt{HG$_{12}^{\star}$}\xspace}
\newcommand{\hgg}{\texttt{HG$_1$G$_2$}\xspace}
\newcommand{\ggparams}{\texttt{G$_1$G$_2$}\xspace}
\newcommand{\gone}{\texttt{G$_1$}\xspace}
\newcommand{\gtwo}{\texttt{G$_2$}\xspace}
\newcommand{\shgg}{\texttt{sHG$_1$G$_2$}\xspace}
\newcommand{\shg}{\texttt{sHG}\xspace}

\newcommand{\filtu}{\ensuremath{u}\xspace}
\newcommand{\filtg}{\ensuremath{g}\xspace}
\newcommand{\filtr}{\ensuremath{r}\xspace}
\newcommand{\filti}{\ensuremath{i}\xspace}
\newcommand{\filtz}{\ensuremath{z}\xspace}
\newcommand{\filty}{\ensuremath{y}\xspace}

\newcommand{\colorgr}{\filtg-\filtr}

\newcommand{\coloriz}{\filti-\filtz}

\newcommand{\Hg}{\ensuremath{H_g}\xspace}
\newcommand{\Hr}{\ensuremath{H_r}\xspace}

\newcommand{\colorHgr}{\Hg-\Hr}

\newcommand{\ztfAlertPerNightReceived}{215,000\xspace}

\newcommand{\ztfTotalObsSSO}{19,319,067\xspace}

\newcommand{\finkTotObjSSO}{565,045\xspace}

\newcommand{\finkTotObjFilteredSSO}{122,675\xspace}

\newcommand{\finkObs}{13,245,908\xspace}

\newcommand{\finkSampleInter}{63,092\xspace}
\newcommand{\finkSuccessInter}{51\%\xspace}
\newcommand{\finkSampleUnion}{95,593\xspace}
\newcommand{\finkSuccessUnion}{78\%\xspace}

\newcommand{\finkSuccessHGGInter}{38\%\xspace}

\newcommand{\finkSuccessHGGUnion}{70\%\xspace}

\newcommand{\finkSuccessHGUnion}{98\%\xspace}

\newcommand{\nObs}{$92_{-26}^{+46}$}
\newcommand{\nDay}{$1,164_{-108}^{+270}$}
\newcommand{\damitSSO}{10,743\xspace}
\newcommand{\damitModel}{16,317\xspace}
\newcommand{\spinComparison}{6,499\xspace}

\usepackage[acronym]{glossaries}
\newacronym{sso}{SSO}{Solar System Object}
\newacronym{LSST}{LSST}{Legacy Survey of Space and Time}

\makeglossaries

\begin{document}

\newcommand{\BC}[1]{\textcolor{purple}{[#1]}}
\newcommand{\JP}[1]{\textcolor{cyan}{[#1]}}
\newcommand{\MM}[1]{\textcolor{Apricot}{[#1]}}
\newcommand{\JB}[1]{\textcolor{green}{[#1]}}
\newcommand{\RLM}[1]{\textcolor{pink}{[#1]}}

\newcommand{\rev}[1]{#1}

\ifx\destination\arxiv
\title{Combined spin orientation and phase function of asteroids}

\author[1]{B. Carry}
\author[2]{J. Peloton}
\author[2]{R. Le Montagner}
\author[3]{M. Mahlke}
\author[4]{J. Berthier}

\address[1]{Universit{\'e} C{\^o}te d'Azur, Observatoire de la C{\^o}te d'Azur, CNRS, Laboratoire Lagrange, Nice, France}
\address[2]{Universit{\'e} Paris-Saclay, CNRS/IN2P3, IJCLab, 91405 Orsay, France}
\address[3]{Institut d'Astrophysique Spatiale, Université Paris-Saclay, CNRS, F-91405 Orsay, France}
\address[4]{IMCCE, Observatoire de Paris, PSL Research University, CNRS, Sorbonne Universit{\'e}s, UPMC Univ Paris 06, Univ. Lille, Paris, France}

  \twocolumn[
    \begin{@twocolumnfalse}
      \maketitle
      \begin{abstract}
   Large sky surveys provide numerous non-targeted observations of
   small bodies of the Solar System. The upcoming \gls{LSST}
   of the Vera C. Rubin observatory will be the largest
   source of small body photometry in the next decade. With non-coordinated
   epochs of observation, colors, and therefore taxonomy and composition, can
   only be computed by comparing absolute magnitudes obtained in
   each filter by solving the phase function (evolution of brightness
   of the small body against the solar phase angle).
   Current models in use in the community (\hg, \hgs, \hgg)
   however fail to reproduce the long-term photometry of many targets
   due to the change in aspect angle between apparitions.
   We aim at deriving a generic yet simple phase function model
   accounting for the variable geometry of the small bodies over multiple
   apparitions. 
   As a spinoff of the \hgg model, we propose the \shgg phase function model
   in which we introduce a term describing the brightness changes due to
   spin orientation and polar oblateness.
   We apply this new model to \numb{\finkObs} observations of
   \numb{\finkTotObjFilteredSSO} SSOs. These observations
   were acquired in the \filtg and \filtr filters with the
   Zwicky Transient Facility between 2019/11/01 and 2023/12/01.
   We retrieve them and implement the new \shgg model in
   \fink, a broker of alerts designed for the LSST.
   The \shgg model leads to smaller residuals than other phase function models,
   providing a better description of the photometry of asteroids. 
   We determine the absolute magnitude $H$ and phase function coefficients 
   (\gone, \gtwo) in each filter, the spin orientation ($\alpha_0$, $\delta_0$), and
   the polar-to-equatorial oblateness $R$ for \numb{\finkSampleUnion} \glspl{sso},
   which constitutes about a tenfold increase
   in the number of characterised objects compared to current
   census.
   The application of the \shgg model on ZTF alert data using the FINK broker shows 
   that the model is appropriate to extract physical properties of asteroids 
   from multi-band and sparse photometry,
   such as the forthcoming LSST survey. 
 abstract

      \end{abstract}
    \end{@twocolumnfalse}
  ]
\newcommand{\degr}{\ensuremath{^{\textrm{o}}}}
\fi

\ifx\destination\aanda
  \title{Combined spin orientation and phase function of asteroids}
  \subtitle{}
  \titlerunning{}
  \authorrunning{Carry et al.}

  \author{%
    B.~Carry\inst{\ref{oca}}  \and
    J.~Peloton\inst{\ref{icjlab}}   \and
    R.~Le Montagner\inst{\ref{icjlab}} \and
    M.~Mahlke\inst{\ref{ias}}  \and
    J.~Berthier\inst{\ref{imcce}}
    }

  \institute{
    Universit{\'e} C{\^o}te d'Azur, Observatoire de la C{\^o}te d'Azur, CNRS, Laboratoire Lagrange, France
    \label{oca}
    \and
    Universit{\'e} Paris-Saclay, CNRS/IN2P3, IJCLab, 91405 Orsay, France
    \label{icjlab}
    \and
    Institut d'Astrophysique Spatiale, Université Paris-Saclay, CNRS, F-91405 Orsay, France
    \label{ias}
    \and
    IMCCE, Observatoire de Paris, PSL Research University, CNRS, Sorbonne Universit{\'e}s, UPMC Univ Paris 06, Univ. Lille, Paris, France
    \label{imcce}
  }

  \date{Received date / Accepted date --
   }
   \abstract
   {Large sky surveys provide numerous non-targeted observations of
   small bodies of the Solar System. The upcoming \gls{LSST}
   of the Vera C. Rubin observatory will be the largest
   source of small body photometry in the next decade. With non-coordinated
   epochs of observation, colors, and therefore taxonomy and composition, can
   only be computed by comparing absolute magnitudes obtained in
   each filter by solving the phase function (evolution of brightness
   of the small body against the solar phase angle).
   Current models in use in the community (\hg, \hgs, \hgg)
   however fail to reproduce the long-term photometry of many targets due to the change in aspect angle between apparitions.
   }
   {We aim at deriving a generic yet simple phase function model
   accounting for the variable geometry of the small bodies over multiple
   apparitions. }
   {As a spinoff of the \hgg model, we propose the \shgg phase function model
   in which we introduce a term describing the brightness changes due to
   spin orientation and polar oblateness.
   We apply this new model to \numb{\finkObs} observations of
   \numb{\finkTotObjFilteredSSO} SSOs. These observations
   were acquired in the \filtg and \filtr filters with the
   Zwicky Transient Facility between 2019/11/01 and 2023/12/01.
   We retrieve them and implement the new \shgg model in
   \fink, a broker of alerts designed for the LSST.}
   {
   The \shgg model leads to smaller residuals than other phase function models,
   providing a better description of the photometry of asteroids. 
   We determine the absolute magnitude $H$ and phase function coefficients 
   (\gone, \gtwo) in each filter, the spin orientation ($\alpha_0$, $\delta_0$), and
   the polar-to-equatorial oblateness $R$ for \numb{\finkSampleUnion} \glspl{sso},
   which constitutes about a tenfold increase
   in the number of characterised objects compared to current
   census.
   }
   {
   The application of the \shgg model on ZTF alert data using the FINK broker shows 
   that the model is appropriate to extract physical properties of asteroids 
   from multi-band and sparse photometry,
   such as the forthcoming LSST survey. 
   }

  \maketitle

\fi



\section{Introduction}

Called the vermins of the sky by astronomers for a long time owing
to the trails they left on photographic plates
\rev{\citep{1930PASP...42....5S}}, the accidental observations
of asteroids in large sky survey have seen a growing interest.
While the cadence and mode of operation of most surveys are seldom optimized for
moving objects \citep{2014AN....335..142S}, the tremendous amount of data
acquired in modern times can provide a wealth of information on
the compositional and
physical properties of \glspl{sso}, unattainable by dedicated observations and
yet crucial to decipher the events that sculpted
our Solar System.

Spectrophotometry is required to determine the taxonomy
\citep{2013Icar..226..723D, 2018A&A...617A..12P},
hence composition, of objects and to map their distribution,
relics of the timing and place of formation, together with
past dynamical events \citep{2014Natur.505..629D}.
Photometry time-series are necessary to determine the
rotation period and spin coordinates
\citep{2001Icar..153...37K, 2004A&A...422L..39K}, which are critical parameters
dictating the dynamical evolution of asteroids through the
Yarkovsky effect \citep{1998Icar..132..378F, 2015-AsteroidsIV-Vokrouhlicky}, spreading
dynamical structures over time \citep{2001Sci...294.1693B, 2006Icar..182..118V}.

Decades of targeted observations
\citep[e.g.,][]{
1985Icar...61..355Z,
1995Icar..115....1X,
2002Icar..158..146B,
2004Icar..172..179L,
2014Icar..229..392D,
2017Icar..291..268L,
2018Icar..304...31D,
2018Icar..311...35D,
2019Icar..324...41B} have, however, brought about
\numb{7,000} visible and
near-infrared spectra only (see the compilation
in the \citet{2022A&A...665A..26M} taxonomy).
The situation is even more dramatic for physical properties, focusing
here on rotation period and spin-axis coordinates.
Detailed studies using stellar occultations, disk-resolved imaging,
or radar echoes
have characterized a few tens of \glspl{sso} only
\citep[e.g.][]{
2000Sci...289.2088V,
2006Sci...314.1276O,
2010Icar..205..460C,
2011Sci...334..487S,
2015MNRAS.448.3382T,
2018Icar..309..134P,
2021A&A...654A..56V}.
The most-productive method has been lightcurve inversion, mainly
using the inversion algorithm of \citet{2001Icar..153...37K}.
However, decades of patient accumulation of lightcurves only brought
solutions for a few hundred \glspl{sso}
\citep{2002Icar..159..369K,
2003Icar..164..346T,
2003Icar..162..285S,
2007A&A...465..331D}.

The game-changer has been the data mining of
serendipitous observations of \glspl{sso} in large sky surveys,
through dedicated software.
The Sloan Digital Sky Survey (\sdss),
the ESO VISTA Hemispherical Survey (VHS),
and the \sm Southern Survey (SMSS) have brought
hundreds of thousands of multi-filter photometry of asteroids
\citep[e.g.,][]{2001AJ....122.2749I, 2016Icar..268..340C, 2016A&A...591A.115P,
2022A&A...658A.109S},
resulting in the determination of the taxonomic class of about
\numb{143,000} \glspl{sso} \citep[\numb{11}\% of the population,
see the compilation in the \ssodnet service\footnote{\url{https://ssp.imcce.fr/webservices/ssodnet}},][]{2023A&A...671A.151B}.
The ESA \gaia mission released
spectroscopy for over \numb{60,000} asteroids \citep{2022arXiv220612174G, 2023MNRAS.519.2917O, 2023A&A...671A..40G}.
The Catalina Sky Survey,
the Lowell Observatory database,
the NASA \kepler and \tess,
the Palomar Transient Factory (PTF), and ESA \gaia
have brought a wealth of photometry on most targets
\citep{2013A&A...559A.134H,
2015AJ....150...75W,
2015ApJS..219...27C,
2016MNRAS.458.3394B,
2016A&A...587A..48D,2019A&A...631A...2D,2023A&A...675A..24D,
2020ApJS..247...26P,
2018A&A...620A..91D,
2018A&A...616A..13G,
2020A&A...642A.138M,
2021ApJS..254....7K},
albeit mainly
sparse in time (i.e., the frequency of observations is much lower
than the intrinsic rotation frequency).
Most important is the theoretical framework to benefit from
this photometry sparse in time \citep{2004A&A...422L..39K}.
Neverthess, rotation periods are available for about
\numb{33,000} asteroids only (\numb{2.5}\% of the population)
and spin coordinates for
\numb{10,000} (\numb{0.7}\%) only \citep[see][]{2023A&A...671A.151B}.

%


The upcoming \gls{LSST} of the
Vera C. Rubin observatory is expected
to discover around \numb{5 million} \glspl{sso} \citep{2009-Book-LSST}.
The determination
of their taxonomy and surface composition based on the provided observations is not simple.
Owing to their irregular shape and constantly changing geometry, the
photometry of \glspl{sso} is a degenerate combination of intrinsic
color and geometry. The determination of their colors hence generally
relies on photometry acquired over short timescales, typically within minutes
\citep{2016A&A...591A.115P}.
While this condition is fulfilled by the modes of operation of the \sdss, \sm,
and the ESA \euclid mission \citep{2018A&A...609A.113C,
2021A&A...652A..59S, 2022A&A...658A.109S},
the \gls{LSST} will provide sparse photometry only \citep{2009EMP..105..101J}.
The determination of the colors of \glspl{sso} must therefore rely on the determination
of their absolute magnitude in each filter
\citep{2021Icar..35414094M, 2022A&A...667A..81A}.
It is been repeatedly shown, however, that the absolute magnitude
\rev{and the slope of the phase function may vary from apparition to apparition
\citep[e.g.,][]{1992LIACo..30..353K,
  2015A&A...580A..98C, 
  2021Icar..35414094M,
  2022MNRAS.513.3076J}.}
This severely increases the required number of observations per \glspl{sso} in a given
apparition to determine its phase curve as observations from different
apparitions cannot be combined as of today.

We introduce here a new model to benefit from
the sparse photometry obtained
over multiple bands from large sky surveys. We aim to provide
a more accurate description of the photometry by simultaneously determining
the absolute magnitude and phase coefficients in
each bands as well as the base geometric properties
(spin coordinates and oblateness) of the \glspl{sso}.
We describe this new model for multi-filter sparse photometry
in \cref{sec:model}. We present
in \cref{sec:data} the data we use
to validate the approach by
comparing the new model with previous results
from the community in \cref{sec:valid}.
We present results on asteroid phase function and
spin orientation in \cref{sec:results}. We finally provide an
easy and programmatic access to these results
in \cref{sec:access}.

\section{Generalized model for sparse photometry\label{sec:model}}

The absolute magnitude $H$ of an \gls{sso}
is defined as the magnitude of the object located at
a heliocentric distance $\Delta$ of \SI{1}{au},
a range to observer $r$ of \SI{1}{au},
and a phase angle $\gamma$ of \ang{0}

\begin{eqnarray}
H &=& H(r=1,\Delta=1,\gamma=0) \nonumber\\
  &=& m - \funcf - \funcg\label[equation]{eq:H}
\end{eqnarray}

\noindent where
\begin{eqnarray}
\funcf &=& 5 \log_{10}(r\Delta) \\
\funcg &=& - 2.5\log_{10}\left[ \gone \phi_1(\gamma) \right.\nonumber \\
  && \hspace{3.1em} +   \gtwo \phi_2(\gamma)\nonumber \\
  && \hspace{3.1em}\left. + ( 1 - \gone -\gtwo) \phi_3(\gamma) \right]
\end{eqnarray}

\noindent with the conditions
\begin{subequations}
\label{eq:gg}
\begin{align}
	0 &<\gone, \label{eq:gg_a} \\ 
	0 &<\gtwo, \label{eq:gg_b} \\
	0 &<1- \gone - \gtwo \label{eq:gg_c}
\end{align}
\end{subequations}

\noindent following the \hgg model by \citet{2010Icar..209..542M}, which was accepted in
2012 by the International Astronomical Union (IAU) to supersede the historical \hg model \citep{1989-AsteroidsII-Bowell}.
The \hgg model offers a better description of the surge of brightness
at low phase angles (called the opposition effect), hence providing
a more accurate determination of the absolute magnitude
\citep{2010Icar..209..542M, 2021Icar..35414094M}.
Furthermore, the \ggparams parameters have been shown to be linked with
albedo and taxonomic type \citep{2016P&SS..123..101S, 2021Icar..35414094M}.

Although the \hgg model presents improvements over the \hg model
both in prediction and interpretation,
its usage has remained limited so far.
It has been used in a few studies, mainly based on
targeted observations
\citep{2019A&A...626A..87S,
2021P&SS..20205248S,
2022A&A...666A.190S}
but also from sky surveys
\citep[in particular ESA \gaia,][]{2021A&A...649A..98M,2021MNRAS.504..761C},
including an initial extensive study by \citet{2011-JQSRT-112-Oszkiewicz}.
However,
ephemerides computation centers (Minor Planet
Center\footnote{\url{https://minorplanetcenter.net}},
Jet Propulsion Laboratory Solar System
Dynamics\footnote{\url{https://ssd.jpl.nasa.gov}}, and
Lowell Observatory \astorb\footnote{\url{https://asteroid.lowell.edu/}})
report absolute magnitudes with the \hg model and not \hgg.
Pragmatically, while \hg always converges
(especially as $G$ is almost always fixed to the canonical value of 0.15),
\hgg has strong requirements on phase coverage to
produce meaningful results \citep[in particular requiring
observations at low phase angles, typically below
2--4\degr, see][]{2021Icar..35414094M}.
A two-parameters version of \hgg was proposed for that purpose
by \citet{2010Icar..209..542M} and refined by \citet{2016P&SS..123..117P},
\hgs, although the absolute magnitudes derived with this latter model
present systematic discrepancies with \hgg
\citep{2021Icar..35414094M} due to the restricted parameter space.

However, the parameters of \hgg are wavelength dependent.
First, the absolute magnitude $H$ is expected to be
different for each observing band
owing to the intrinsic color of the asteroid
\citep[close to Solar colors as a first approximation, see][]{2021-riea}.
This is already the case for the \hg model,
in which $H$ is implicitly $H_V$, the
absolute magnitude reported in Johnson V
band\footnote{\url{http://svo2.cab.inta-csic.es/theory/fps/index.php?id=Generic/Johnson.V}
\citep{2012ivoa.rept.1015R}}.
The advantage of measuring the color based on absolute
magnitudes is to avoid biases introduced by differences in
observing time and brightness variation related to the
shape of the  \gls{sso}
\citep{2016A&A...591A.115P, 2018A&A...609A.113C, 2022A&A...667A..81A}.
Second, the \ggparams are also different \citep{2021Icar..35414094M},
something which has been reported as
\textsl{phase reddening} and observed spectrally on
both laboratory samples and asteroids in the sky
\citep{2012Icar..220...36S, 2019Icar..324...41B, 2024arXiv240211113A}.

Furthermore,
the \hgg model suffers from a major limitation:
it does not account for the non-spherical geometry of asteroids
\citep{2022MNRAS.513.3076J}. Owing to the
changing aspect angle (the angle between the spin axis and the
viewing direction) over apparitions, the absolute magnitude can
differ \citep[see Fig.~10 in][for instance]{2021Icar..35414094M}.
Conversely, a single \hg or \hgg fit on a multi-opposition data
set will provide an average absolute magnitude but will likely result
in biased magnitude predictions.

In the shape modeling formalism introduced by
\citet{2001Icar..153...37K}, this aspect is solved by
the simultaneous modeling of the 3D shape and the phase
function (with a linear-exponential model). It, however,
typically requires more data (either dense light curves or
many photometry measurements sparse in time) than for
phase-function fitting only
\citep[see the discussion in][]{2015-AsteroidsIV-Durech, 2023A&A...675A..24D}.

We propose here an intermediate solution,
dubbed \shgg for \texttt{spinned}\hgg, solving the
apparition-to-apparition change of magnitude without
adding too many requirements on the data set.
\rev{It is in essence the 
``reference phase curve'' defined by \citet{2001Icar..153...37K}.}
We introduce a new term $\funcs$ to the definition of the absolute
magnitude $H$ (\cref{eq:H}), accounting for the orientation of the asteroid
and function of its
oblateness $R$
and the equatorial coordinates of
its spin axis ($\alpha_0$, $\delta_0$):
\begin{eqnarray}
  H &=& m - \funcf - \funcg - \funcs \label[equation]{eq:SH},
\end{eqnarray}

\begin{figure}[t]
  \centering
  \input{gfx/fig_model.pgf}
  \caption{%
    The \shgg model of (22) Kalliope, with its
    three components --
    \funcf,
    \funcg,
    \funcs\ --
    shown explicitly, together with their amplitude.
    The \hgg model is plotted for comparison.
    }
  \label{fig:model}
\end{figure}

\noindent where
\begin{equation}
  \funcs = 2.5 \log_{10} \Big[ 1 - (1 - R) |\cos \Lambda| \Big],
\end{equation}

\noindent with $\Lambda$ the aspect angle, computed from the equatorial coordinates
($\alpha$,$\delta$) of the asteroid at the time of observation as
\begin{equation}
  \cos \Lambda = \sin{\delta} \sin{\delta_0} + \cos{\delta}\cos{\delta_0}\cos{(\alpha - \alpha_0)}.
\end{equation}

Thus, an additional criterion is added to the
definition of $H$: it corresponds to the magnitude
of the object at \SI{1}{au} from both the Sun and the observer,
with a \ang{0} phase angle, \textsl{seen from its equatorial plane}
As a corollary, the object is brighter by $2.5 \log_{10} R$ magnitudes
when observed from its pole.
Noting $a>b>c$ the tri-axial diameters of the asteroid,
$R$ is the polar-to-equatorial oblateness ($0 < R \le 1$):
\begin{equation}
  R = \frac{c(a+b)}{2ab} \label[equation]{eq:R}
\end{equation}

\begin{figure}[t]
  \centering
  \input{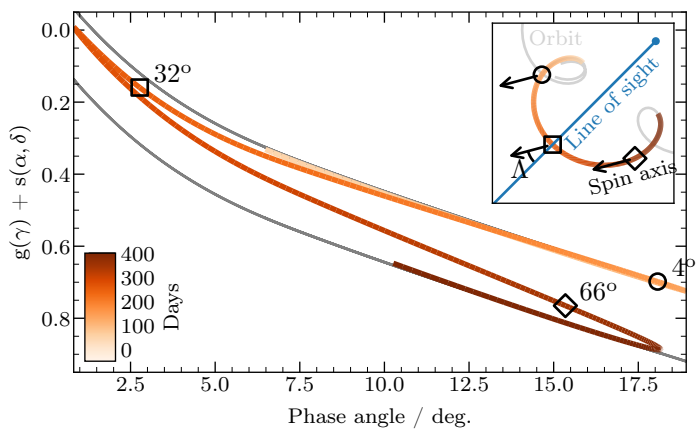}
  \caption{%
    Effect of the changing geometry (described by
    \funcs) on the phase curve
    (traditionally \funcg only).
    Over 400 days from JD 2,460,000 (same as
    \cref{fig:model}), (22) Kalliope is seen from
    an aspect angle increasing from 4\degr
    (pole on) to 78\degr (close to the equator)
    owing to motion as seen from the Earth.
    The inset presents the successive
    ecliptic ($x$,$y$) positions of Kalliope,
    color-coded by epoch, in
    a reference frame centered on Earth
    (the pale blue dot).
    Three reference epochs are drawn on both graphs
    (circle, square, and diamond).
    The orientation of the spin axis of Kalliope
    is drawn as arrows at these three epochs, and the
    line of sight is drawn for the second (square) epoch,
    illustrating the aspect angle $\Lambda$.
    }
  \label{fig:phase}
\end{figure}

Applying this new definition of $H$ on observations taken over
several apparitions with $N_f$ different filters,
one can simultaneously determine
the absolute magnitude $H$ and phase parameters \ggparams for each
band and the oblateness
and spin coordinates
for a total of $3 \times (N_f + 1)$ parameters.
We present in \cref{fig:model}
an illustration of the \shgg model
for the asteroid (22) Kalliope
\citep[we use the spin solution from][]{2022A&A...662A..71F}.
The distance term $\funcf$ presents the expected
minima and maxima at oppositions and conjunctions.
The phase term \funcg has a periodicity twice faster, with
decreasing phase angle at both oppositions and conjunctions. Both
are symmetric around epochs of minimal phase angle (as visible
on the blue and orange curves). Their sum defines the \hgg model
\citep[][]{2010Icar..209..542M}.
The new term \funcs accounts for the slow change of geometry
over time, allowing asymmetry.
We present in \cref{fig:phase} the phase and spin components
(\funcg and \funcs) in a similar fashion
to Fig.~5 of \citet{2022MNRAS.513.3076J}.
The phase curve is bounded between two extreme cases,
pole on ($\Lambda = \{0\degr, 180\degr\}$) and
equator on ($\Lambda = 90\degr$),
and practically presents a slow evolution between these
boundaries.

\section{Data\label{sec:data}}

We implement the \shgg model in
\fink\footnote{\url{https://fink-broker.org/}}
\citep{2021MNRAS.501.3272M},
a broker of alerts for the \gls{LSST}
\citep{2009-Book-LSST}.
Today, before the start of the \gls{LSST},
\fink processes daily the public stream of alerts from the
Zwicky Transient Survey
\citep[\ztf,][]{2019PASP..131a8003M, 2019PASP..131g8001G,
2019PASP..131a8002B, 2019PASP..131a8001P}.

\ztf broadcasts a public stream for variable and transient events in two bands
\citep[\filtg\footnote{\url{http://svo2.cab.inta-csic.es/theory/fps/index.php?id=Palomar/ZTF.g}} and
  \filtr\footnote{\url{http://svo2.cab.inta-csic.es/theory/fps/index.php?id=Palomar/ZTF.r}},
  similar but not identical to the
  Sloan \filtg and \filtr filters,][]{2019PASP..131a8002B}
of typically
\numb{\ztfAlertPerNightReceived} alerts per night, of which about \numb{70}\% is retained by \fink for scientific analyses, and
\numb{14}\% correspond to \glspl{sso}. Between late \numb{2019} and late \numb{2023}, \fink has extracted
\numb{\ztfTotalObsSSO} observations of \glspl{sso} from the \ztf public alert stream.
\fink is one of the seven brokers of alerts being developed for the
\gls{LSST}. As such, the system has been designed to cope with a very large
flow of data, on all aspects of astrophysics, from Solar system to
variable stars, supernovae, and optical counterparts to gravitational wave events
\citep[e.g.,][]{2021MNRAS.501.3272M, 2022MNRAS.515.6007A, 2022A&A...663A..13L, finkfat}.
With the \shgg model implemented within
\fink, its results are already freely available
for each \gls{sso} through the \fink Web interface\footnote{\url{https://fink-portal.org/}}
and the corresponding Application Public Interface (API) \apisso. We also propose a dedicated method
to retrieve the parameters of the \shgg model for a large corpus of data at once:
the Solar System Objects \fink Table\footnote{\url{https://fink-portal.org/api}} (\ssoft,
see \cref{sec:access}).

We use \fink to retrieve
\numb{\ztfTotalObsSSO} observations of
\numb{\finkTotObjSSO} unique \glspl{sso} in \filtg and \filtr between 2019/11 and 2023/12.
We decide to only retain objects with at least \numb{50} observations across all filter bands
(\numb{\finkTotObjFilteredSSO} unique \glspl{sso}).
\rev{This threshold in number of observations was chosen arbitrarily
after several tests, as a compromise between the sample size and the 
fraction of objects failing to converge to a solution (see Sect.~\ref{sec:valid:fail}).}
The average number of observations (together with the 25\%
and 75\% quantiles) is \numb{\nObs}, for a length of
observations (number of days between the first and the last
observation) of \numb{\nDay}.
We fit these observations with the
\hg \citep{1989-AsteroidsII-Bowell},
\hgg \citep{2010Icar..209..542M}, and
\shgg models, and compare the results in the following
sections.

\begin{figure*}[t]
  \centering
  \input{gfx/223_vs_time.pgf}
  \caption{%
    Comparison of the \hg, \hgg, and \shgg models on the
    photometry of \nuna{223}{Rosa} from \ztf.
    }
  \label{fig:example}
\end{figure*}

\section{Validation of the \shgg model\label{sec:valid}}

  The improvement provided by \shgg is qualitatively visible
  on \cref{fig:example} for the asteroid \nuna{223}{Rosa}, a
  flyby candidate by the ESA \juice mission
  \citep{2013P&SS...78....1G, 2022P&SS..21605476A}.
  While the description of the opposition effect is different
  between \hg and \hgg, they both predict a globally symetric behavior
  around opposition.
  \shgg correctly describes the photometry as it
  accounts for the evolving geometry with time.
  We use the \numb{\finkTotObjFilteredSSO} \glspl{sso} to assert statistically the
  validity and limits of the \shgg model.

\subsection{Success and failure\label{sec:valid:fail}}

  \rev{We impose a set of boundaries for the parameters in the 
  fitting process to guarantee meaningful results. Especially we
  enforce that the conditions in \Cref{eq:gg_a} and \Cref{eq:gg_b}
  are fulfilled, and that the oblateness $R$ is in between 0.3 and 1
  (encompassing all published shape models, see Sect.~\ref{sec:valid:spin})}.

  \rev{Out of the \numb{\finkTotObjFilteredSSO} 
  initial \glspl{sso} lightcurves,
  the fitting procedure for the \shgg model converges in about
  \numb{98}\% of the cases. However at this stage, there are suspicious
  solutions, where the minimization algorithm is clearly hitting 
  the boundary conditions. Hence in the following, we consider a solution 
  fully valid if: the fitting procedure converges, 
  the $G_1$ and $G_2$ values are non singular 
  (singular cases encompass values stricly equal 0), 
  the spin coordinates are non singular 
  (singular cases encompass $\alpha_0$ strictly equals to 
  0\degr~or 360\degr, ($\alpha_0$,$\beta_0$) equal (180\degr,0\degr)), 
  and the condition in \Cref{eq:gg_c} is fulfilled
  (it could not be incorporated in the fitting procedure).}

  Using the two filter bands of ZTF ($g$ and $r$), this results in a sample
  of \numb{\finkSampleInter} \glspl{sso} 
  (\numb{\finkSuccessInter} of success).
  This means about
  half of the fitted parameters are of good quality in all bands
  simultaneously \citep[a similar success rate as][on \gaia
  photometry]{2023A&A...675A..24D}.
  As the phase parameters are fitted per band,
  we also define the sample of \glspl{sso} for which the
  constraints on spin parameters defined above are fulfilled, and \gone and
  \gtwo fulfill the conditions in \Cref{eq:gg} for at least one filter band.
  This results in a sample of \numb{\finkSampleUnion} \glspl{sso} (\numb{\finkSuccessUnion} of success).
  This sample will be used for the rest of the analysis unless explicitly stated.

  For the \hg model, the rate of success is
  higher as expected from its simplistic form (\numb{\finkSuccessHGUnion} of success), but the \hgg model has somewhat
  a lower success rate than the \shgg model (\numb{\finkSuccessHGGInter} of success in the two filter bands simultaneously,
  and \numb{\finkSuccessHGGUnion} of success in at least one filter band).
  This gives confidence that the spin component to the lightcurve model contributes
  to ameliorate the parameters estimation globally compared to the \hgg model.

  Nonetheless, we have tried to understand the reasons for failure for the \hgg and \shgg models.
  The failures are not clearly correlated with the minimum phase angle, the number of observations per filter,
  the number of oppositions, and the range for the aspect angle values.
  Based on \cref{fig:model}, we see that the contribution of \funcg and \funcs
  components are relatively small compared to the other components,
  and often smaller than the typical error estimate on the photometric
  measurements reported in the \ztf alert packets.
  Hence, we might be in a noisy regime where the contribution from
  the phase and spin components are not always accessible.
  To test this hypothesis, we take the valid solutions for which
  the fitted procedure converges, and we
  randomize the magnitudes for each observation with a Gaussian
  distribution within (a) the reported observational errors,
  and (b) 10 times the reported observational errors. We fit again
  the models parameters, and restart the procedure 500 times.
  In the (a) scenario, the results remain stable, within the reported standard deviations.
  On the contrary the (b) scenario systematically leads to outlier
  values for the \hgg and \shgg models similarly to what is observed on the initial dataset.
  Although this is perhaps not the only reason, we are confident
  that the rate of success should improve with higher signal-to-noise measurements,
  such as the data that will be soon collected by the Vera C. Rubin Observatory.

\subsection{Fit to data\label{sec:valid:rms}}

\begin{figure}[t]
  \centering
  \input{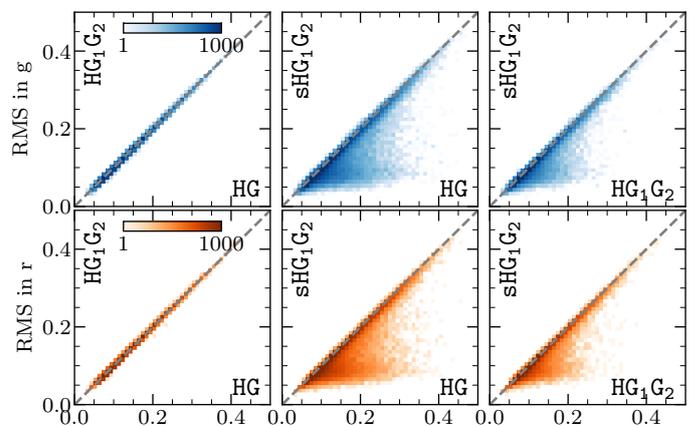}
  \caption{%
    Comparison of the residuals for the
    \hg, \hgg, \shgg models in both
    \filtg and \filtr filters.
    }
  \label{fig:rms}
\end{figure}

\begin{figure*}[t]
  \centering
  \input{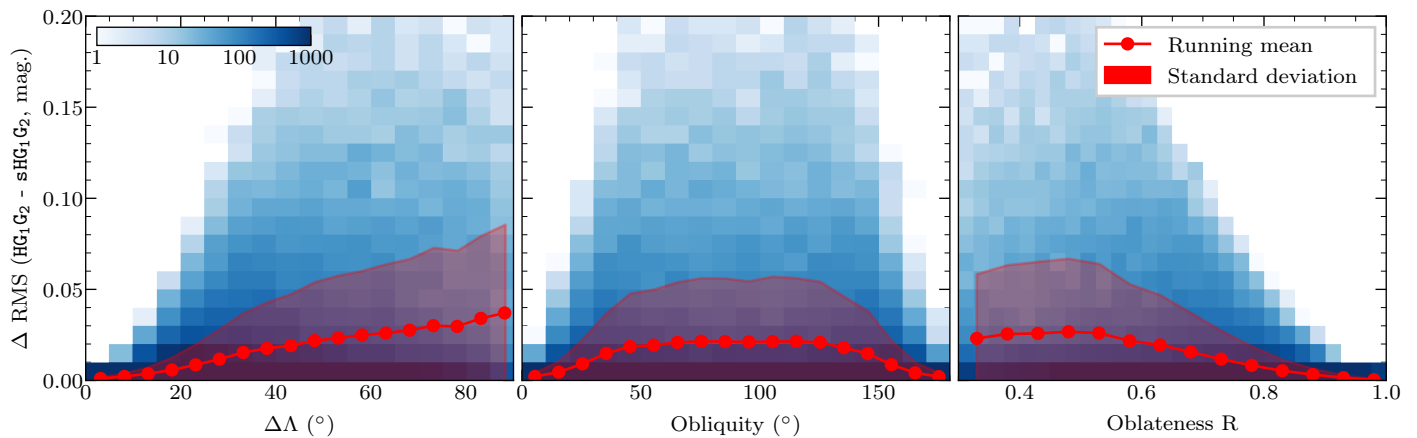}
  \caption{%
    Comparison of the residuals for \hgg and \shgg models as
    function of the range of aspect angle ($\Delta \Lambda$),
    the obliquity, and the oblateness $R$.
    Larger values of $\Delta$RMS indicate cases where
    \shgg improves over \hgg.
    }
  \label{fig:rms_limits}
\end{figure*}

We compare the root mean square (RMS) residuals between the
predicted photometry by \hg, \hgg, and \shgg models and
the \ztf photometry in \filtg and \filtr in
\cref{fig:rms}.
The improvement of \hgg over \hg
(\cref{fig:rms}, left) is marginal, with all \glspl{sso}
close to the diagonal. The improvement of \shgg over \hgg
(\cref{fig:rms}, right)
is clear, and it is striking over \hg (\cref{fig:rms}, middle),
with a significant tail below the diagonal.

The bulk of the population remains close to the diagonal.
This implies that the predictions by \hg or \hgg fit
equally well the data as \shgg, although the latter has more
degrees of freedom.
There are three cases in which \shgg may not be required.
First, if the \gls{sso} is (nearly) spherical, its orientation will
not change significantly its brightness. The \funcs term of \shgg
is therefore not required and converges toward zero.
Second, if obliquity of the \gls{sso} is close to either
\ang{0} or \ang{180}
\citep[as expected from YORP evolution, and as it is often the
case, see \cref{sec:valid:spin} and][]{2015-AsteroidsIV-Vokrouhlicky,
2015-AsteroidsIV-Durech, 2023A&A...675A..24D},
as the orbital inclination is small the range of aspect
angle $\Lambda$ covered remains limited around \ang{90}. The
brightness hence barely changes over apparitions, and the
shape/spin term \funcs contribution is minimal.
Third, depending on the time span (mainly the number of
apparitions) and number of observations, the range of aspect
angle $\Lambda$ may again be limited.
These three limiting cases are visible on
\cref{fig:rms_limits}.

Two conclusions can be drawn from this comparison of RMS.
First, the \shgg model indeed provides a clear improvement over the
previous models. Second, this improvement is not always needed,
for the reasons listed here above.
Pragmatically, this implies that \hg, \hgg, and \shgg models should
be computed for each \gls{sso} (in \fink but more generally at any
ephemerides computation center): from the most robust to the most
informative. In the pathological cases listed above, \hgg should
then be preferred over \shgg. Similarly, if low phase angles are not
covered, the two-parameters \ggparams function \funcg (common
to \hgg and \shgg) is not constrained \citep[see][]{2021Icar..35414094M}
and \hg should be preferred.

\subsection{Absolute magnitude\label{sec:valid:H}}

\begin{figure}[t]
  \centering
  \input{gfx/abs_mag.pgf}
  \caption{%
    Comparison of the absolute magnitude in \shgg against
    \hg and \hgg, in \filtg and \filtr filters.
    }
  \label{fig:abs_mag}
\end{figure}

\begin{figure}[t]
  \centering
  \input{gfx/color_comparison.pgf}
  \caption{%
    The \colorgr color from
    \shgg absolute magnitudes (\Hg and \Hr)
    compared with
    \sdss and \sm apparent \colorgr colors
    (all converted to \ps photometric system, see text).
    }
  \label{fig:valid:colors}
\end{figure}

We compare in \cref{fig:abs_mag} the \shgg absolute magnitude
in \filtg and \filtr with those obtained with \hg and \hgg.
The three models are in good agreement,
as shown by the mode of 0 in their difference in
absolute magnitude. The dispersion between \shgg and \hgg is
much smaller than with \hg, owing to the shared definition
of the phase function \funcg between the two models that
differs from \hg.
The two-parameters (\ggparams) function \funcg was introduced by
\citet{2010Icar..209..542M} to better describe the opposition
effect at small phase angles. It is therefore not unexpected
to have differences of absolute magnitude between the two
systems.
There is a clear asymmetry in absolute magnitude,
as expected from the definition
of \shgg (\cref{eq:H}): the absolute magnitude is defined in the
equatorial plane of the \glspl{sso} and is hence larger on average
(the projection area on the plane of the sky is maximal for
objects seen pole on).

As an additional validation, we compare in
\cref{fig:valid:colors} the \ztf \colorgr obtained
from the \Hg and \Hr absolute magnitudes with the \shgg model
with the \colorgr color from the \sdss and \sm surveys
\citep{2021A&A...652A..59S, 2022A&A...658A.109S}.
The \filtg and \filtr of the three facilities differ, and
colors must therefore be corrected before comparison.
We convert all three systems to the \ps \colorgr color, using the
following transformations:
\sm to \sdss \citep{2022A&A...658A.109S},
\sdss to \ps \citep{2016ApJ...822...66F},
and \ztf to \ps \citep{2020RNAAS...4...38M}.
The distributions show a general trend towards the 1:1 relation as expected.
The spread of values is larger for \ztf than for \sdss and \sm, due to the fact that
the \ztf magnitudes have an added uncertainty due to the computation via the phase curve,
while \sdss and \sm can compute the colors directly from the near-simultaneously acquired photometry.
Two separate regions of higher density are visible, corresponding
to the C and S taxonomy complexes (centered on \colorgr values of 0.4 and 0.55,
respectively).
This validates
the approach of color determination from absolute magnitudes with
the \shgg model.

\subsection{Phase parameters \ggparams\label{sec:valid:GG}}

\begin{figure}[t]
  \centering
  \input{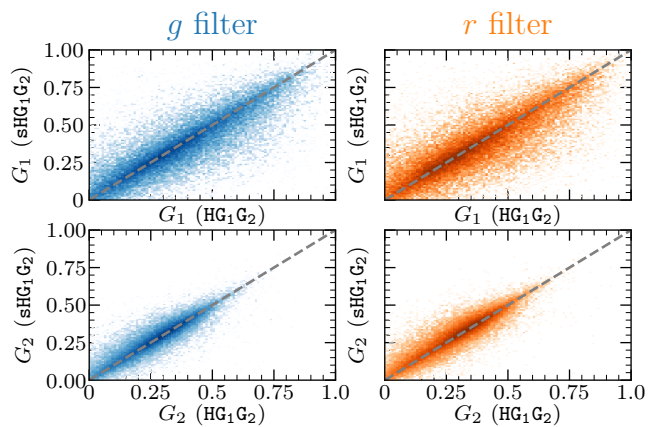}
  \caption{%
    Comparison of the of \ggparams parameters in $g$ and $r$ filters
    for the \hgg (x axis)
	and \shgg (y axis) models.
    }
  \label{fig:g1g2:comp}
\end{figure}

\begin{figure}[t]
  \centering
  \input{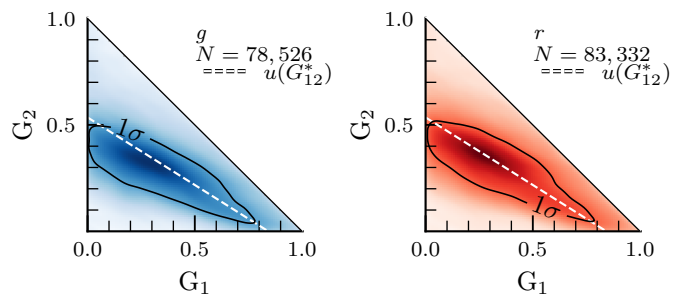}
  \caption{%
    Distribution of \ggparams parameters in $g$ and $r$ for the \shgg model.
    }
  \label{fig:g1g2:distrib}
\end{figure}

  We compare in \cref{fig:g1g2:comp} the values of \ggparams obtained
  with the \shgg model with those from the \hgg model. Both
  $\gone$ and $\gtwo$ show a good agreement, albeit some spread is present.
  The addition of the geometry term \funcs in the definition of the
  absolute magnitude (\cref{eq:SH}) thus does not bias the slope
  parameters \ggparams.
  We further test the values of these parameters in \cref{fig:g1g2:distrib}.
  For both \filtg and \filtr filters, the bulk of the \glspl{sso}
  follow the \hgs line in the $\{\gone,\gtwo\}$ space
  \citep{2010Icar..209..542M,2016P&SS..123..117P,
    2016P&SS..123..101S}, again
  confirming the validity of the \shgg model.

\subsection{Spin coordinates\label{sec:valid:spin}}

\begin{figure}[t]
  \centering
  \input{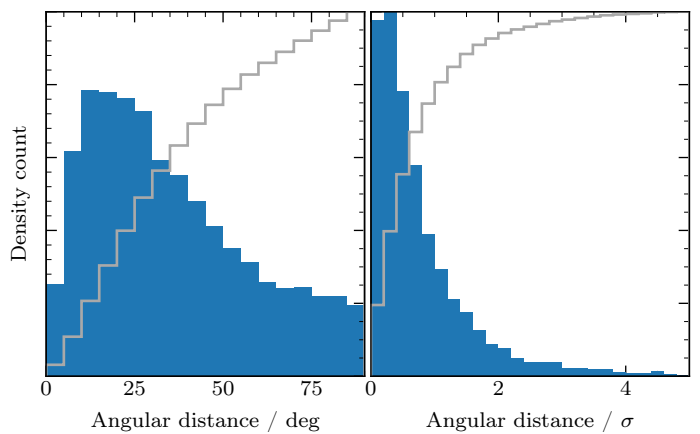}
  \caption{%
    Angular distance between spin-axis coordinates from
    \shgg and literature for
    \numb{\spinComparison} asteroids, in degree (left)
    and normalized by uncertainty (right).
    The cumulative distributions up to 100\%
    are also presented (in grey)
    }
  \label{fig:valid:spin}
\end{figure}

We compare the spin-axis coordinates ($\alpha_0$,$\delta_0$)
derived with the \shgg from those available in the literature.
We collect available spin solutions for the
\numb{\finkSampleUnion} \glspl{sso} in the \fink sample using the
\ssodnet service \citep{2023A&A...671A.151B}.
We find \numb{\spinComparison} \glspl{sso} with previous estimates
of the spin-axis coordinates, mainly from the lightcurve inversion technique
\citep{2001Icar..153...37K} or in combination with
other techniques \citep[stellar occultations, direct imaging,
and thermophysical modeling,][]{2010Icar..205..460C,
2011-IPI-5-Kaasalainen,
2011Icar..214..652D,
2015A&A...576A...8V,
2018Icar..309..297H}
by many authors
\citep[e.g.,][]{2013A&A...551A..67H,
2021A&A...654A..87M,
2021A&A...654A..56V,
2022PSJ.....3...56H},
compiled on the
DAMIT\footnote{\url{https://astro.troja.mff.cuni.cz/projects/damit/}}
service \citep{2010A&A...513A..46D}.

We present in \cref{fig:valid:spin} the distribution of angular
separation between the spin-axis coordinates from the
literature and \shgg.
The distribution peaks around \ang{20}, with half of the \glspl{sso}
agreeing below \ang{35}. This is an overall good agreement,
especially considering the limited number of observations
available to the \shgg model (\numb{\nObs}, see
\cref{sec:data}).
We also present this distribution normalized by the
spin uncertainty (computed as the quadratic sum of the uncertainties
from the \shgg and literature spin coordinates).
About 60\% of the solutions agree within one $\sigma$, and
95\% at three $\sigma$, an indication that uncertainties seems
properly estimated (and not over- or under-estimated).

\subsection{Polar oblateness\label{sec:valid:R}}

\begin{figure}[t]
  \centering
  \input{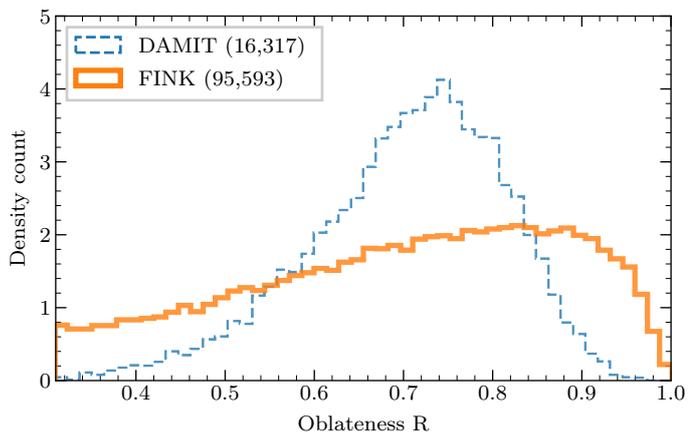}
  \caption{%
    Distribution of the oblateness $R$, compared
    with that of \numb{\damitModel} shape models
    of \numb{\damitSSO} asteroids from
    DAMIT \citep{2010A&A...513A..46D}.
    }
  \label{fig:R}
\end{figure}

\begin{figure}[t]
  \centering
  \input{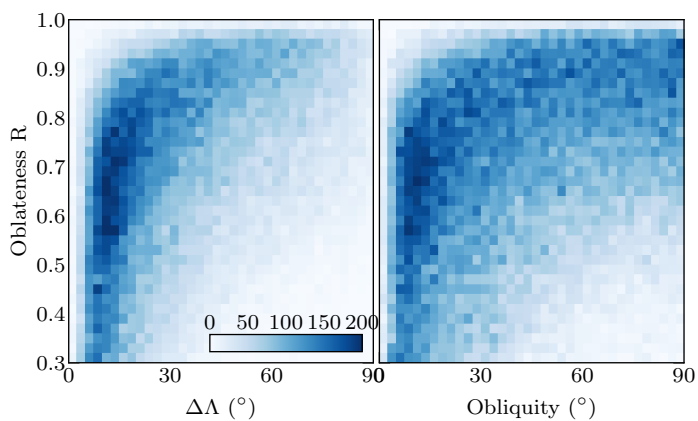}
  \caption{%
    Oblateness $R$ as function of the range
    of covered aspect angles ($\Delta \Lambda$)
    and obliquity (given the symmetry of 
    \funcs, we only plot obliquity in 0-90\degr).
    }
  \label{fig:R_trend}
\end{figure}

We present the
distribution of the polar oblateness $R$ in \cref{fig:R},
compared with that of existing 3-D shape models from
the \rev{scientific} literature
\rev{(a comparison with popular literature
is presented in \Cref{app:clarke})}.
For that, we downloaded the \numb{\damitModel} models
of \numb{\damitSSO} unique asteroids available on
DAMIT
\citep{2010A&A...513A..46D}.
These models are the product of inversion techniques
\citep[mainly][]{
2001Icar..153...24K, 2001Icar..153...37K,
2010Icar..205..460C,
2015A&A...576A...8V} from
many authors
\citep[e.g.][]{
2016A&A...587A..48D,
2018A&A...617A..57D,
2018Icar..299...84H,
2021A&A...654A..48H,
2017A&A...607A.117V}.
We compute the oblateness $R$ of all these models
from Eq.~\ref{eq:R}, computing first the tri-axial
diameters ($a \geq b \geq c$) from the moments of inertia
\citep[using the formulae by][]{1996Icar..124..698D}.

The distribution of oblateness determined here presents two
main differences with the one from the 3-D shape models: it peaks
at rounder objects ($R$ of 0.9 rather than 0.7)
and has a long tail toward flying saucers (low $R$).
The tail is an artifact as most \glspl{sso} with an oblateness
below 0.5 have observations covering a very limited range of
aspect angle and/or an obliquity close to either 0\degr or
180\degr, limiting the viewing geometry as shown in \cref{fig:R_trend}.

The shift in the center of the distribution is more complex to
interpret and could be real or artificial.
While the range of absolute magnitude between the \fink sample and
the shape models from DAMIT overlap, the median absolute magnitude
is \numb{14.8} against \numb{13.6}. The \glspl{sso} reported
here have thus a smaller diameter (with half typically smaller than 4\,km).
The distribution of oblateness between the two samples could genuinely
be different.
On the other hand, the difference aligns with intrinsic biases of both
methods at play.
As shown by \citet{2018A&A...610A...7M} there is a clear bias against
slow rotations and low-amplitude light curves among asteroids with
a determined spin axis and 3-D shape, even more marked as objects get fainter.
The light curve inversion indeed favors targets with a strong intrinsic variability,
which creates a bias against round asteroids (as $a>b>c$, large $a/b$ ratios
imply smaller $R$).
This is confirmed as asteroids with a 3-D shape model have generally larger residuals
with \shgg in \ztf data.
\rev{Therefore, the distribution of oblateness for well-constrained
solutions (e.g., large enough $\Delta \Lambda$, valid \ggparams,
low residuals) is likely representative of the real distribution of
oblateness among asteroids. 
We plan to combine several data sets from different facilities, such as 
\ztf, \atlas, \ps, \gaia, LSST, to increase the time coverage, 
hence range of observed aspect angle $\Lambda$, to further
constrain the obliquity and oblateness distributions.}

\section{Results\label{sec:results}}

\begin{figure}[ht]
  \centering
  \input{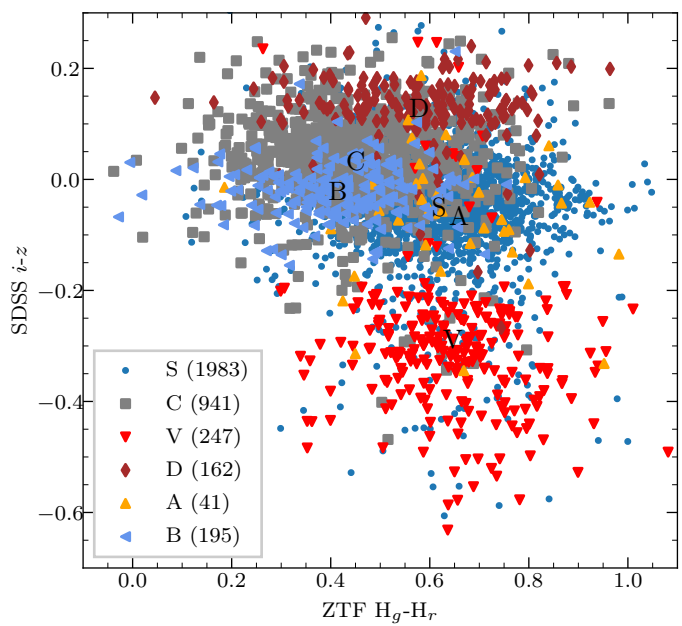}
  \caption{%
    Distribution of \numb{3,569} \glspl{sso} in
    \sdss \coloriz
    against
    \ztf \colorHgr.
    The symbols and colors correspond to the taxonomic
    class retrieved from \ssodnet \citep{2023A&A...671A.151B},
    and the letter mark the average color for each class.
    }
  \label{fig:result:griz}
\end{figure}

  One of the main motivation to develop \shgg was the determination
  of reliable colors from sparse photometry of \glspl{sso}, such as the
  \gls{LSST} will provide. We show in \cref{sec:valid:H} the validity
  of the approach.
  The data set we use here (\ztf) only contains two filters
  (\filtg and \filtr) but it is enough as a proof of concept.
  We illustrate further the approach in \cref{fig:result:griz},
  comparing \ztf \colorHgr with \sdss \coloriz, mimicking
  the situation with \gls{LSST} observations with absolute magnitudes
  determined with \shgg in all six \gls{LSST} filters:
  \filtu,
  \filtg,
  \filtr,
  \filti,
  \filtz, and
  \filty \citep{2009-Book-LSST}.
  The different taxonomic groups are easily identified, and
  follow their expected location in such a ``slope''
  (\colorHgr) versus ``one micron band'' (\coloriz)
  plane \citep{2001AJ....122.2749I,
    2005Icar..173..132N,
    2009Icar..202..160D,
    2013Icar..226..723D,
    2023A&A...679A.148S}.

  This is extremely promising for \gls{LSST}: the \ggparams of the phase function
  are better constrained with larger phase coverage
  \citep{2021Icar..35414094M}, and as a result the absolute magnitude $H$ too.
  Furthermore, the longer
  the time spanned by the observation, the more geometries observed, and
  the better the constraints on the spin-axis and polar oblateness.
  With a factor of four to five more observations (400 \textsl{vs} \numb{\nObs})
  over a time interval three times longer
  (10 years \textsl{vs} three), the colors obtained with
  the \gls{LSST} will be much more precise (not to mention a better
  intrinsic photometric accuracy).

\begin{figure*}[t]
  \centering
  \input{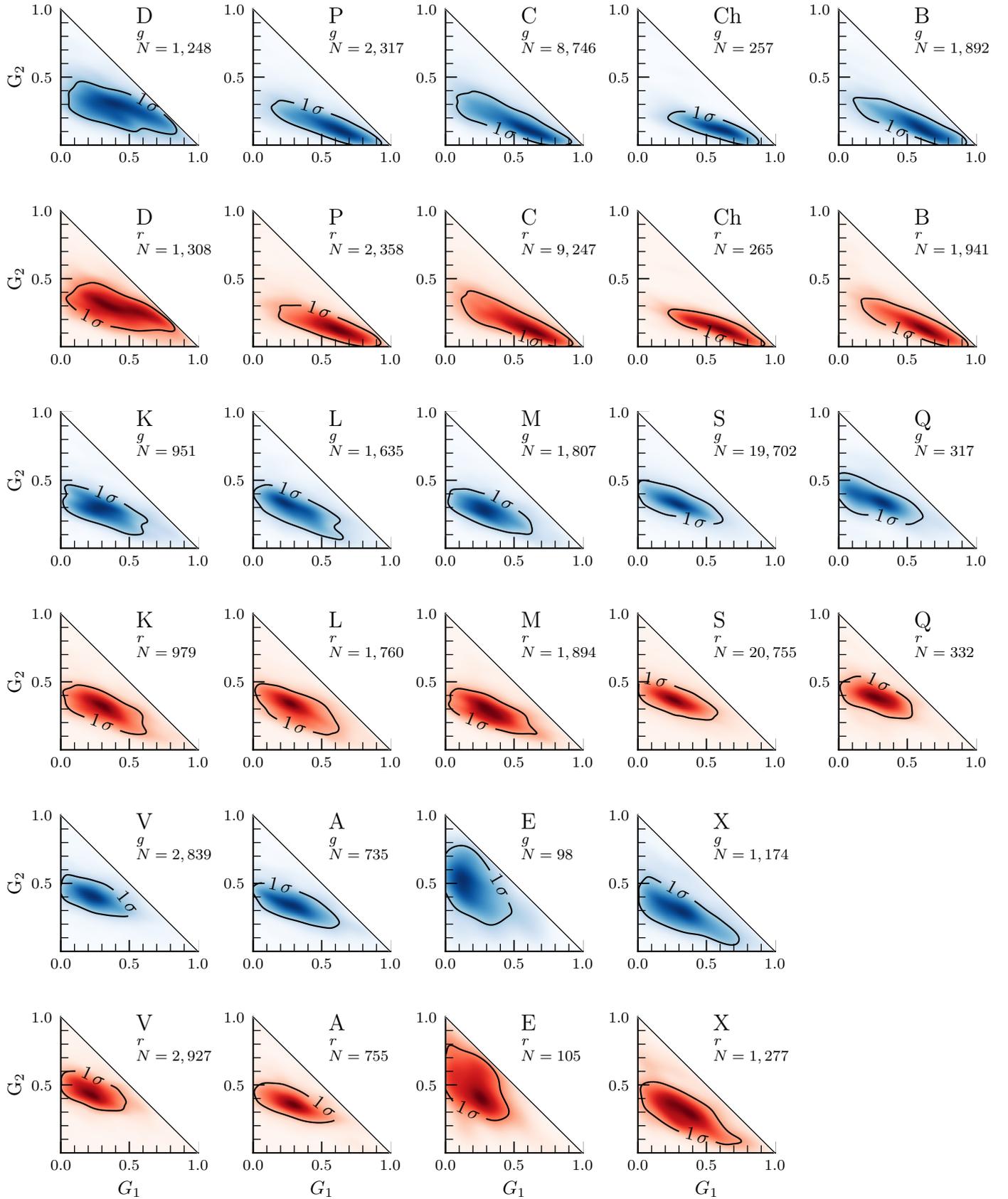}
  \caption{%
    Distribution of \ggparams parameters in $g$ and $r$ per taxonomic complex.
    }
  \label{fig:g1g2}
\end{figure*}

  We then present in \cref{fig:g1g2} the distribution of the \ggparams parameters
  for \filtg and \filtr filters separatedly, for asteroids in the
  taxonomic complexes A, B, C, Ch, D, E, K, L, M, P, Q, S, V and X (i.e.,
  spectrally similar to E/M/P without albedo information), following
  the work by \citet{2021Icar..35414094M}.
  There is a clear trend in \ggparams from low-albedo asteroids
  (C/P/D)
  occupying the high-\gone-low-\gtwo region to the
  high-albedo asteroids (V/A/E) located toward low-\gone-high-\gtwo,
  as expected \citep{2016P&SS..123..101S}.
  The dispersion of each taxonomic group in \cref{fig:g1g2} is more limited
  here than in \citet{2021Icar..35414094M}: it may be due to
  an intrinsic higher precision on \ztf photometry or the \shgg model
  being more adapted, or both. In any case, the strong correlation of
  \ggparams with taxonomy opens the possibility to use these parameters
  in the determination of asteroid taxonomic classes, especially with
  \gls{LSST} photometry.

\begin{figure}[ht]
  \centering
  \input{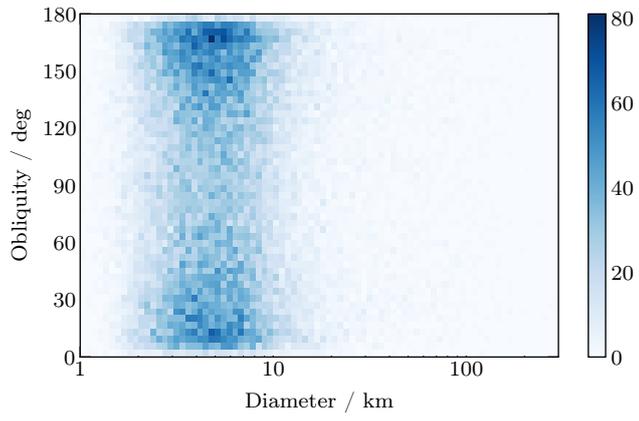}
  \caption{%
    Obliquity as function of the diameter.
    }
  \label{fig:obli}
\end{figure}

\begin{figure*}[ht]
  \centering
  \input{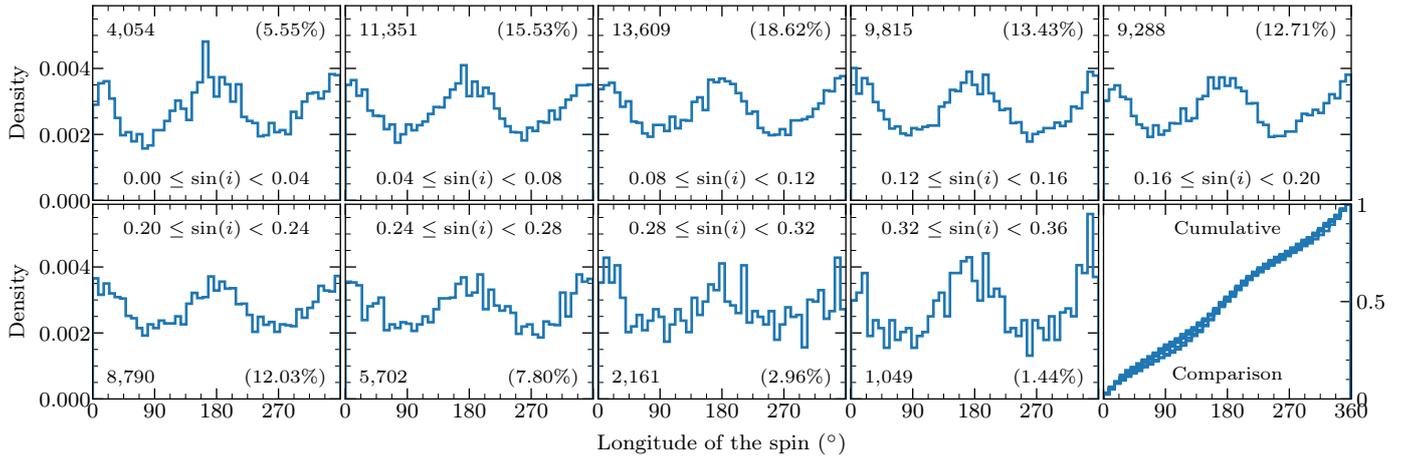}
  \caption{%
    Distribution of the ecliptic longitude
    of the spin axis as function of the orbital inclination.}
  \label{fig:lon_inc}
\end{figure*}

  We compare the obliquity (computed from the spin-coordinates
  determinated with \shgg) with diameter
  \citep[retrieved from \ssodnet,][]{2023A&A...671A.151B}
  in \cref{fig:obli}.
  Asteroids larger than \SI{50}{km}
  appear roughly isotropic
  \citep[although an excess of direct rotators has been
  observed,][]{2010MNRAS.404..475J, 2020Icar..33513380V}
  Asteroids smaller than 10--\SI{20}{km}
  cluster toward extreme values
  of \ang{0} and \ang{180}, although there are many asteroids
  with intermediate obliquity.
  This is a clear print of the
  Yarkovsky–O'Keefe–Radzievskii–Paddack
  effect
  \citep{1952AZh....29..162R,
  1969JGR....74.4379P,
  1976tto..book.....O}, or YORP for short
  \citep{2015-AsteroidsIV-Vokrouhlicky}.
  This obliquity-diameter distribution has been known
  for years
  \citep[e.g.,][]{2013A&A...551A..67H, 2015-AsteroidsIV-Durech}
  The sample of \numb{\finkSampleUnion} from \ztf/\shgg, however,
  increases the sample of known obliquity
  by a factor of about \numb{ten}.

  The next logical step would be to study the distribution of
  obliquities among asteroids belonging to families, such as
  recently done by  \citet{2023A&A...675A..24D}
  However, the \shgg model is by construction symmetric accross
  the equator of the target asteroid, and
  the determination of the spin-axis is thus ambiguous:
  the rotation can be either
  direct or retrograde.
  We hence report two spin solutions for each object in Fink:
  ($\alpha_0$,$\delta_0$) and
  ($\alpha_0$+180,-$\delta_0$).

  We also study the distribution of ecliptic longitude of the pole.
  There is a strong correlation with the longitude of the ascending node, as
  expect from a majority of \glspl{sso} with an obliquity close to
  either 0\degr or 180\degr: the longitude should be close to the
  longitude of the node minus 90\degr.
  We do not find a dependence between
  the distribution of longitude of the pole
  and the orbital inclination (\cref{fig:lon_inc}), as opposed
  to \citet{2016A&A...596A..57C}
  who reported a more isotropic distribution of longitude
  for more inclined orbits.
  We compare the cumulative distributions of longitude for each
  range of inclination in \cref{fig:lon_inc}. They are barely
  distinguishable from one another, which is confirmed by
  a Kolmogorov–Smirnov test: all the distributions are
  statistically similar.

\section{Data availability}\label{sec:access}

  As mentioned in \Cref{sec:data}, the data and results presented
  in the present study were all acquired and processed within
  \fink. The amount of \glspl{sso} for which a solution is determined,
  and the values of the parameters of the \shgg model, are therefore
  in constant evolution.
  \fink processes the incoming stream of alerts from \ztf on a daily basis,
  and the \glspl{sso} parameters are determined once a month.
  The data and derived parameters reported here correspond
  to the \ssoft of \numb{December 2023}.

  As an official community broker of alerts for \gls{LSST} \fink will receive in
  real time the flow of alerts from Vera C. Rubin observatory.
  As such, the absolute magnitude $H$ and \ggparams in each of
  \gls{LSST} filters together with spin coordinates ($\alpha_0$,
  $\delta_0$) and polar oblateness ($R$) will be regularly determined and in open access
  to the scientific community for every \gls{sso} observed in the \gls{LSST}.

  The data and parameters can be retrieved from
  \fink Science Portal and API, in particular the summary table
  \ssoft containing all parameters. In the Python programming language,
  data can be retrieved as
  follows\footnote{More information at \url{https://fink-portal.org/api}}:

  \begin{lstlisting}[language=Python, frame=single, caption={}, label=listing:tracklet]
import requests

r = requests.post(
  "https://fink-portal.org/api/v1/ssoft",
  json={
    "flavor": "SHG1G2",
    "version": "2023.12",
    "output-format": "json"
  }
)
\end{lstlisting}

\section{Conclusion\label{sec:conclu}}

  We propose a simple modification of the currently accepted
  \hgg model to properly render the photometric behavior of
  \glspl{sso} taken over long periods of time.
  From observations in $N_f$ filters, the new \shgg model simultaneously determines
  the absolute magnitude $H$ and phase function coefficient \ggparams in
  each filter,
  the spin coordinates ($\alpha_0$,$\delta_0$),
  and the polar oblateness $R$ for a total
  of $3 \times (N_f + 1)$ parameters.
  The determination of the absolute magnitude accross multiple filters is required
  to determine the colors of \glspl{sso} in large sky surveys without back-to-back
  observations in different filters, such as the upcoming \gls{LSST} of the Vera C. Rubin
  observatory.

  We test the new \shgg model on observations in \filtg and \filtr,
  collected over three years by the \ztf.
  The \shgg model provides a better description of the photometry,
  as revealed by smaller residuals, than previous proposed models
  (\hg and \hgg). The parameters determined by \shgg are nevertheless
  consistent with previous estimates. The absolute magnitude and
  phase parameters are in agreement with those determine with the \hgg
  model on the same data set, and the spin coordinates agree with
  those determined with other methods on different data sets within
  typically 20--30\degr.
  The polar oblateness present a spurious trend toward small values, and
  a systematic shift in its peak value compared to \glspl{sso} with shape
  model, which could be genuine but is also aligned with the expected
  biases from the different methods at play.

  The limitations of the \shgg model are linked with the geometry of observations.
  Observations at low phase angle (between 0\degr~and 5\degr) are required
  to properly describe the opposition effect (affecting the \ggparams parameters),
  a limitation inherited from the \hgg model.
  The spin orientation and oblateness cannot be properly retrieved from observations
  spanning only a limited range of geometries, in particular the aspect angle.
  Finally, by construction of the model, there is a complete degeneracy between
  direct and prograde rotation.
  \rev{The new term \funcs accounting for the oblateness and orientation can,
   however, also be introduced in the \hg model, in a \shg model.
  Practically, all four \hg, \shg, \hgg, and \shgg models should be computed for
  each \gls{sso}. }
  Then,
  following Occam's razor principle,
  the simpliest model, among those fitting the data satisfactorily, should
  be chosen and used.

  The \shgg model is fully implemented in \fink, a broker of alerts for
  the \gls{LSST}. The results of the model are freely available on \fink portal,
  for all \glspl{sso} currently observed by \ztf for which the fitting procedure succeeds,
  and by the \gls{LSST} in a nearby future.

\section*{Acknowledgements}%
\label{sec:acknowledgements}%

This project has received financial support from the CNRS
through the MITI interdisciplinary program.
B.C. was supported by CNRS/INSU/PNP.
We thank these programs for their support.
This research used the
\miriade \citep{2009-EPSC-Berthier},
\ssodnet \citep{2023A&A...671A.151B},
and
\topcat
\citep{2005ASPC..347...29T}
Virtual Observatory tools.
It used the
\astropy
\citep{astropy:2013, astropy:2018, astropy:2022},
\sbpy
\citep{2019JOSS....4.1426M}, and
\rocks \citep{2023A&A...671A.151B}
python packages. This work was developed within the \fink
community and made use of the \fink community broker resources.
\fink is supported by LSST-France and CNRS/IN2P3.
Thanks to all the developers and maintainers.

\clearpage

\ifx\destination\arxiv
  \bibliographystyle{aux/arxiv}
\fi

\ifx\destination\aanda
 \bibliographystyle{aux/aa} 
\fi

\ifx\destination\publisher
  \bibliographystyle{aux/publisher}
  \biboptions{authoryear}
\fi

\bibliography{aux/bib}


\appendix
\section{Good bet, Mr Clarke!\label{app:clarke}}

\say{\textit{On Day 86 they were due to make their closest approach to any known
asteroid, It had no name - merely the number 7794 [...]
Through the high-powered telescope, they could see that the asteroid was
very irregular, and turning slowly end over end. Sometimes it looked like a
flattened sphere, sometimes it resembled a roughly shaped block; [...]}} \textit{ - Arthur C. Clarke, 2001 -- A Space Odyssey} \\

In the 1968 Arthur C. Clarke book, \textit{2001 -- A Space Odyssey}, the author describes the observation
of the asteroid 7794 from a flyby (which would be discovered only in 1996\footnote{\url{https://minorplanetcenter.net/db_search/show_object?object_id=7794}}),
predicting that this asteroid \textit{"looked like a flattened sphere"}.
Using the most recent data from \fink (version \texttt{2024.01},
see \cref{sec:access}),
we found a value for the oblateness around 0.5, favouring the prediction made in the book.
The parameter estimate has still a large uncertainty though, but
given the residuals in the phase curve,
we can infer that either the asteroid is irregular along its equator
($a \neq b$) or there could be an extra unmodelled activity.

\end{document}